# David Brewster's and William Herschel's experiments on inflection that delivered the *coup de grâce* to Thomas Young's ether distribution hypothesis

Olivier Morizot

Aix-Marseille Univ, CNRS, Centre Gilles Gaston Granger, Aix-en-Provence, France

**Abstract**
In his *Theory of Light and Colours*, presented to the Royal Society in November 1801, Thomas Young defended a mechanical explanation of the coloured fringes observed outside of the shadow of an opaque object – the so-called 'colours by inflection' – that was based on the hypothesis of an ethereal density gradient surrounding all material bodies. However, two years later, he publicly rejected that hypothesis, without giving much detail on his reasons. Now, although Geoffrey Cantor demonstrated the crucial role of mechanical and astronomical arguments in explaining the withdrawal of this fundamental hypothesis long ago, the purpose of this article is to draw deeper attention on a set of experiments performed by David Brewster on the inflection of light, described in a letter he addressed to the Royal Society in January 1802, but which kept retained in the hands of William Herschel and never reached its original destination. For the hypothesis that will be evaluated here is that these unpublished experiments of Brewster's were eventually known to Young through the mediation of William Herschel, and eventually played a significant role in his rejection of his own ether distribution hypothesis too.

It is well known that in November 1801 Thomas Young presented to the Royal Society a vibratory theory of light based on the four hypotheses that 'a luminiferous ether pervades the universe, rare and elastic in a high degree'; that 'undulations are excited in this ether whenever a body becomes luminous'; that 'the sensation of different colours depends on the frequency of vibrations, excited by light in the retina'; and that 'all material bodies have an attraction for the ethereal medium, by means of which it is accumulated within their substance, and for a small distance around them, in a state of greater density, but not of greater elasticity' (Young 1802a: 14-22).

   This last hypothesis actually consists of two parts: first, that the ethereal medium is denser inside material bodies than it is in air; which is consistent with Young's early assessment that light waves are analogous to longitudinal sound waves and that, as such, their velocity would be independent of their amplitude and frequency, but would decrease with the density of their conveying medium (Young 1802a: 22-23). And which provided him with a simple justification of refraction, based on the deceleration of light passing from air to glass or water.[1] As well as an explanation of partial and total reflection, as a consequence of the partial or total reflux of light waves trying to engage into a medium with different ether density (Young 1800: 127-128; 1802a: 30-31).

---
[1] An explanation that was therefore compatible with those of Huygens (1690) and Euler (1746).

Second, and most important for our matter, that the ether density was continuously decreasing from its value inside the material body to its value in the surrounding medium; thus creating a thin layer of varying ether density above the surface of all solid bodies that, in Young's eyes, could explain the so-called 'colours by inflection' (Young 1802a: 42-44). Indeed, according to him, a portion of light passing very close to a material body would enter this ethereal atmosphere of increasing density. So that its velocity would progressively decrease, and its direction would bend towards the body; up to a point when some light would hit the body and be reflected from its surface in a direction that would depend not only on the incidence angle of the ray at the entrance of the atmosphere, but also – most certainly – on the shape and structure of the surface of the body, and on the radial profile of its ethereal gradient. Those portions of light reflected in different directions would then interfere with other portions that would have passed just far enough from the body not to be deviated, and produce the periodic and coloured pattern observable at the outside edge of the shadow.

Thus, it seems that Young had developed at that time a simple and coherent explanation of the main cases of the deviation of light (refraction, reflection, inflection) based on the single consideration of the mechanical consequences of the local ether density on the propagation of the undulations of light.[2] Moreover, years ago, Geoffrey Cantor showed how this second part of Young's fourth hypothesis – that he named the 'ether distribution hypothesis' – was then a key element in Young's wider project for a unified natural philosophy; reducing all phenomena of light, heat, electricity, magnetism and material cohesion to the dynamics of a single particulate ether (Cantor 1970a).

**Withdrawing the ether distribution hypothesis**

However, in that same paper, Cantor also stressed that Young had certainly abandoned this ether distribution hypothesis by the first half of year 1802 – that is just a few months after the publication of his theory – although this decision would only be printed in his 1803 Bakerian Lecture, where he would publicly state that:

> I have not, in the course of these investigations, found any reason to suppose the presence of such an inflecting medium in the neighbourhood of dense substances as I was formerly inclined to attribute to them; and, upon considering the phenomena of the aberration of the stars, I am disposed to believe, that the luminiferous ether pervades the substance of all material bodies with little or no resistance, as freely perhaps as the wind passes through a grove of trees. (Young 1804: 12-13)

And that this decision led him to adopt another mechanical justification of the inflection of light 'by the tendency of light to diverge'.[3] And consequently, forced him to rewrite completely that fourth hypothesis and its corollary on 'colours by inflection' – and to omit the corollary on 'blackness' – as he reprinted his *Theory of Light and Colours* in 1807 (Young 1807).

---

[2] 'For explaining the phenomena of partial and total reflection, refraction, and inflection, nothing more is necessary than to suppose all refracting media to retain, by their attraction, a greater or less quantity of the luminous ether, so as to make its density greater than that which it possesses in a vacuum, without increasing its elasticity; and that light is a propagation of an impulse communicated to this ether by luminous bodies' (Young 1800: 127). On top of that, Young also thought by that time that one could infer an explanation 'of blackness' from that ether gradient hypothesis (Young 1802a: 42)

[3] Indeed, as one shall see later, Young had already used this plain explanation of the divergence of light waves – that he credited to Hooke – in his *Theory*, within the frame of an argument against the classical objection that light waves should skirt around obstacles and be seen in their shadows (Young 1802a: 25-30; Darrigol 2012: 176-179). And even had suggested it as an alternative mechanism that could be at the origin of the inflection fringes (Young 1802a: 43).

On top of that, Cantor then developed a very justified explanation of the reason why Thomas Young had given up on this distribution hypothesis, mainly based on a thorough analysis the notebooks for his lectures at Royal Institution (Cantor 1970b) and connecting this decision essentially to considerations on the phenomena of cohesion, of the free movement of planets in the ethereal medium and of the aberration of stars. A set of considerations which seem to have developed in Young's mind by summer 1802, and which finally led him to consider the impossibility of reconciling his ether distribution hypothesis with some necessity of accepting that ether could flow freely through material bodies – as suggested by the previous quotation (Young 1804: 12-13).

However, one cannot help being surprised that optical arguments finally occupy so little space in Cantor's interpretation of Young's radical conclusion (Cantor 1970a: 49). All the more so since such arguments were already furnished by Young himself, along with his first optical paper. Indeed, the tenth section of his *Outlines of Experiments and Inquiries respecting Sound and Light*, dedicated to 'the analogy between light and sound', closes with such words: 'It does not appear that any comparative experiments have been made on the inflection of light by substances possessed of different refractive powers; undoubtedly some very interesting conclusions might be expected from the inquiry' (Young 1800: 130).

Such that one feels allowed to interpret his claim – at the end of his 1801 Bakerian Conference – that 'I do not consider it as quite certain, until further experiments have been made on the inflecting power of different substances, that Dr. Hooke's explanation of inflection, by the tendency of light to diverge, may not have some pretensions to truth' (Young 1802a: 43) as the concession that the experimental comparison of the inflection patterns produced by different bodies could be decisive for deciding whether or not 'the tendency of light to diverge' could provide a better explanation of colours by inflection than his own ethereal atmospheres. Indeed, since different bodies – and, particularly, bodies with different refractive powers – would retain different ether densities, which – for sake of continuity – would most necessarily lead to different ether density gradients surrounding them, the independence of inflection patterns on the nature of the inflecting body would be incompatible with the ether gradient hypothesis; but it would still comply with Hooke's hypothesis.

Such that those two public declarations of Young's certainly are intimately connected to another conclusion that can be found in the sixteenth of his notebooks for his lectures at the Royal Institution,[4] most probably dating from the first half of May 1802, and stating that:
> The cause of the inflection of light is not yet sufficiently understood, but there is great reason to suppose that the first opinion respecting it is the most accurate, and that it arises as Dr. Hooke supposes principally from the natural tendency of light to diverge in all directions: for if it were conceived to originate from the same cause that produces refraction, it might be expected to differ in degree when produced by different inflecting substances: this however is not the case. (Young 1802b: XVI, 17r)[5]

---

[4] The twenty notebooks related to the lectures Thomas Young gave at the Royal institution of Great Britain in 1802 and 1803 were consulted in their manuscript version, held at University College London. From now on, the pages of these manuscripts will be referred by the number of the corresponding notebook (as indicated on its cover) in Roman signs, followed by the Arabic number of the folio (n.1 corresponding to the cover of the notebook, on which Young used to write too) and letter 'r' or 'v', that will stand for the recto or verso of the folio. Note that these notebooks were extensively analyzed and commented by Cantor (1970b).

[5] This sentence illustrates Young's concluding lecture on 'Hydrodynamics', on the 17$^{th}$ of May 1802, in which he regrets he had no more than four months to prepare his all set of fifty lectures on natural philosophy. As far as one can say considering these notebooks, they seem to contain mostly the information he was intending to deliver to his audience on the first session of this course of lectures in 1802, in the precise form he was intending to formulate

For this quotation not only confirms that, just a few months after the publication of his *Theory of light*, Young had already rejected the idea that inflection could be produced by a continuous match at the surface of a material body between the two different ether densities – that at its core and that of the surrounding medium – that are the cause of refraction. But for it also suggests that Young came to that conclusion at least partly on the basis of the comparison of the inflections produced by different substances; which he apparently found out to be precisely the same. And the feeling of conviction conveyed by the brevity of the statement, coupled to the inconvenience for Young of its consequences, strongly suggests that he actually obtained a firm proof of it. And, even so the absence of any further consideration on this point – neither in his published texts, nor in the remains of his personal notebooks and correspondence – duly justified Cantor's intuition that Young probably did not perform that 'crucial' experiment (Cantor 1970a: 49), and fully legitimated his thorough inquiry and conclusions, still an intriguing question remains: could not he have obtained such experimental evidence from someone else's work? And that is the question that will motivate the rest of our inquiry.

**Brewster's missing letter to the Royal Society**

As a matter of fact, there is no trace of any similar experiment in the articles published at the time in the journals Young was familiar with. Yet, the archives of the Royal Astronomical Society contain an interesting letter, sent from Edinburgh on 19 January 1802 by David Brewster to David Erskine, the 11$^{th}$ Earl of Buchan, with the request that it be presented to the Royal Society of London. However, the letter was forwarded to William Herschel first, between whose hands it staid for more than a year before he sent it back to Brewster. And although the original letter cannot be found, a copy of it still lies in Herschel's archives (Brewster 1802a).

> This letter, then, is dedicated to the determination of 'the cause of the Phenomena of Inflexion',[6] within the frame of Brewster's projectile conception of light. And indeed, after reviewing the main opinions of his time on the causes of inflection – alluding to those of Grimaldi and Hooke, describing those of Newton, Mairan, Du Tour and Brougham more precisely, and omitting the most recently defended by Thomas Young, that he was probably unaware of – Brewster underlines their inadequacies, and offers to study the case himself in three steps, that are:
>> to enquire whether or not those rays of light which are most refrangible, are least Flexible; – to make some observations calculated to determine the law of the Force which bodies exert, in inflecting the rays of light; – and to show that there is no necessity for supposing the existence of a propelling force in producing the phenomena of inflexion. (Brewster 1802a: 1)

---

that information, and were completed along with the progression of the course. Meaning that this precise comment was most probably written during the week preceding 17 May 1802 – as references in the same lecture to the discovery of the colours of mixed plate he had probably made that same week actually tend to confirm (Young 1802b: XVI, 16v). Yet, it is not impossible that it was added in the first half of year 1803, as Young repeated the lectures for the second and last time. The fact that the content of these notebooks seems uniquely dedicated to be read in front of the audience of the Royal Institution leaves little probability that this sentence was inserted at any other time, although Young still beheld these notebooks as he was editing a printed version of his lectures in 1807 (Young 1807), in which that precise comment does not appear anymore.

[6] Brewster and Erskine write 'inflexion' or 'flexion' what Young or Herschel spell 'inflection' or 'flection'. We will therefore conserve letter 'x' as we will be citing Brewster's or Erskine's own words only, and use the other and more common spelling the rest of the time.

In short, the first section of the investigation starts with an *a priori* theoretical analysis of the problem, based on the 'universally admitted' opinion at the time that light consisted of projectiles emitted from the sun, and that the seven primary colours were produced by particles differing either in size or velocity. Brewster thus draws on the Newtonian model explaining refraction by the exertion on the projectiles of light of a force perpendicular to the surface; and dispersion by a difference in the momenta before refraction of the projectiles producing the different primary colours. This first theoretical analysis is then followed by a complete reinterpretation of Henry Brougham's recent experiments on inflection (Brougham 1796), where Brewster particularly points out that their author had neglected the important fact that the retina should be less sensitive to projectiles producing a violet sensation than to those producing red – which present a greater momentum – when comparing the width of the shadow of small objects inserted in coloured beams of light, leading to an erroneous comparison of the inflections separately obtained in red and violet light. Both parts of this section leading Brewster to the conclusion that 'those rays of light which are most refrangible, are likewise most inflexible' (Brewster 1802a:6).

And the third section of the letter makes use of Gibbes Walker Jordan's latest observations on inflection (Jordan 1799) in order to demonstrate that neither the concept of a repulsive force, nor that of a very short-range attractive force decreasing with distance and turning into a repulsion after it reaches zero – an explanation based on Newton's considerations on the question (Newton 1730: 370-371), and most widespread at the time – are necessary to explain inflection. On the contrary, Brewster concludes, 'there is every reason to believe that nothing but an attractive force is exerted in producing the phenomena of Inflexion' (Brewster 1802a: 12). And we will turn back to this conclusion later.

For the second section of that letter will concern us most now. In fact, Brewster first determines there the law of this attractive inflective force he has postulated: relying on a personal analysis of Newton's experiments on the inflection of light by two crossed knives (Newton 1730: 305-307), he concludes that 'the force, therefore, exerted by the knives upon the light which forms the hyperbolic fringes, appears to be in the simple inverse ratio of the distance' (Brewster 1802a: 8). A conclusion which immediately leads to a series of experiments, aiming at highlighting the differences between this inflective force and the gravitational one: because these two forces had so often been considered as analogous along the previous century, that Brewster foresees that this supposed analogy might be a source of objections to his own result if he does not take immediate care of undermining it.

That is why – after an interesting epistemological argument on the weakness of analogical arguments in natural philosophy – Brewster turns to the first experimental observation that, although the gravitational force 'exerted by any body is proportional to the quantity of mass it contains, or the number of particles it is composed of […]; the point of a pin, or a grain of sand will inflect the rays of light as much from their rectilinear direction, as bodies of the greatest magnitude, and produce the same changes upon the passing rays of light' (Brewster 1802a: 8-9). A first step from which Cantor, again, justly pointed out in his reference work on *Optics after Newton,* that Brewster will later deduce that this inflective force probably resides in the ultimate particles of matter (Cantor 1983: 82).

However, legitimately focused at the time on the comparison of the various projectile theories of flection, Cantor did not draw the attention on the fact that a second set of experiments, that Brewster describes right after, seems precisely to address the case of the experiment imagined by Thomas Young as a decisive test of his ether distribution hypothesis. Brewster writes:

> In order to determine whether or not the different texture of bodies might cause a difference in the phenomena, I held in the conical beam of light, propagated in a dark room, through a small hole in the window-shut, the stamina of different plants, sulphureous bodies, spunge, Flint glass, Iceland Chrystal, and a great variety of other bodies which from the difference of their texture, or from their producing different effects in the refraction of light, might have been expected to bend the rays of light into different angles of deviation; but in all these cases the breadth of the fringes, and their distance from the Inflecting body, were universally the same. (Brewster 1802a: 9)

Obviously, the theoretical frame and the intention of the experiment performed by Brewster are not those imagined by Young. Indeed, after having dismissed the influence of its size, or mass, Brewster now focuses on the potential influence on inflection of the texture or nature of the inflecting body. But incidentally, as he multiplies the observations of the inflection patterns produced by different bodies, he very explicitly reviews the case of a variety of bodies 'producing different effects in the refraction of light', including flint glass and Iceland crystal; and formally concludes that their effect on the inflection of light is 'universally the same'. Clearly, his intention here is to dismiss another set of objections to his model of the inflecting force, based on its disrespect of the analogy with the supposed refractive force, also responsible for dispersion. But, were Thomas Young aware and confident in the result of such experiments, there is little doubt that he would have concluded there would be no more 'reason to suppose the presence of such an inflecting medium in the neighbourhood of dense substances as I was formerly inclined to attribute to them' (Young 1804: 12).[7]

All the more than Brewster's next and final experiment directly compares the inflection fringes produced by a rod of Iceland crystal crossing a rod of an unspecified sulphureous body, which, 'as far as the eye could judge', happen to be exactly the same (Brewster 1802a: 9). And since, in the frame of Young's theory of light, both the double refraction of the Iceland crystal and the sulphureous quality of bodies would necessarily be the consequence of a specific inner ether density – different from that of other common refractive or opaque bodies, and most certainly different from one to the other – it is most probable that, in Young's eyes, this last experiment would not only demonstrate the absence of consequence on inflection of the mass of these two bodies, but also of their inner ether density.[8]

Therefore, as Thomas Young publicly claimed on two public occasions that experiments on the inflecting power of different refractive bodies would offer a test of his ether distribution hypothesis; as there is no trace of such experiments neither in his published or unpublished papers, nor in the literature of that time; yet, as his notebook number XVI (completed some time between January 1802 and May 1803, but most probably in May 1802) provides evidence of his absolute conviction that inflection is finally independent on the refractive power of the inflecting body; and as Brewster's letter was sent precisely within the short period separating Young's rejection of his ether distribution hypothesis for the explanation of inflection from his last use of the ether distribution hypothesis (in his *Theory of Light and Colours* dating November 1801); it is very tempting to forecast that, one way or another, the results of the experiments described in Brewster's letter on inflection came to Thomas Young's attention early in 1802, and played a significant role in his withdrawal of the ether distribution hypothesis – at least from his optical theory.[9] Although on this assumption, the remaining question would still be: how did the information sealed in this letter finally reach Young?

---

[7] Note that the result of the first experimental observation would probably have already been considered by Young as sufficient to dismiss his ether distribution hypothesis, as he claims for instance in his notebooks that 'In general there is considerable analogy between this refractive density and the specific gravity of the substance. [...] In order of refractive density, beginning from the lowest, or a vacuum, we have rarer air, denser air with various elastic fluids, water, which is the least refractive of all liquids, spirits, oils, glass, and lastly the diamond; but probably some metallic substances are still more refractive' (Young 1802b: XV, 12r).
[8] See for instance: 'in general inflammable bodies are more refractive than bodies of the same density not inflammable' (Young 1802b: XV, 12r).

For, although it was destined to the Royal Society – and would have then ended public – the letter, as mentioned earlier, never reached this destination.[10] And, as far as one can tell, its content was only known to David Brewster himself; to David Steuart Erskine, 11th Earl of Buchan, to whom he entrusted it; and to William Herschel, to whom the letter was forwarded, but who retained it for more than a year before returning it to sender.[11]

**The missing link**

Although we do not have a definite answer to this question, it is still possible to point to some interesting leads. And first, to discard as very unlikely the hypothesis that Thomas Young was informed of the results of these experiments by David Erskine. Although he had been a Fellow of the Royal Society since 1765, and although his younger brother Henry David Erskine, later 12th Earl of Buchan, was to marry Thomas Young's sister-in-law Caroline Maxwell in 1839 (ten years after Thomas' death), there is little evidence of any further connection between those two men. And Erskine seems very loosely connected to the problem at stake anyways, as testified by the two very brief letters of his, dating from this period and concerning that matter, that still can be found in Herschel's correspondence: the first, dated 8 March 1802, mentions his worry that 'a paquet containing a paper on the Inflexion of the rays of Light by Mr. Brewster of Edinburgh may have been miscarried on the road to Slough having been submitted to Dr. Herschel's consideration with the approbation of the author, it was transmitted on the 15 of January' (Erskine 1802). And the tone and content of that letter not only reveal the Earl's eagerness to receive Herschel's confirmation that the letter reached him – and to see it promptly transmitted to the Society. But also, the fact that it was transferred almost immediately from Brewster to Herschel,[12] Erskine merely serving as a relay between them. Then the second letter, sent a year later, simply demands that Brewster's letter be returned to its author as soon as possible (Erskine 1803).

Concerning David Brewster himself, he was only nineteen at the time and not a member of the Royal Society yet – which explains why the letter was entrusted to Erskine, and then to Herschel, to be communicated to that assembly. He did not know Thomas Young personally, and probably did not even know his works in optics at the time he wrote the letter. Still, in a series of notes published in the July 1802 edition of the *Edinburgh Magazine*, Brewster offers a brief digest of Young's *Theory of Light*, which he explicitly and negatively connects, a page later, to the case of the hyperbolic inflection fringes that was central to his own letter:

> There is one insuperable objection to the undulatory system to which its reviver Dr. Young paid no attention. This system cannot explain the hyperbolic fringes which Newton mentions in the 10$^{th}$ Observation of the third book of his Optics, or, rather the very existence of these fringes is an irrefragable argument in favour of the hypothesis which Newton framed. (Brewster 1802b: 64)

---

[9] Since Cantor actually observed that from that period, because of dissonant arguments coming from several different branches of natural philosophy, Young not only gave up on his ether distribution hypothesis, but also on his ambition to construct a complete system of natural philosophy based on the actions of a single ethereal fluid (Cantor 1970a: 61-62), the absence of direct optical evidence of the inexistence of such atmospheres of luminiferous ether itself could indeed have allowed him to preserve this model in the strict case of optics.

[10] As confirmed by the minutes of the meetings at the Royal Society for 1802 and 1803 (Royal Society of London 1799; 1803), the letter was neither read, nor publicly mentioned, in that institution.

[11] Note that William's sister, Caroline Herschel, was then probably aware of that letter too, as the version of it that can now be found in the Herschel Archives at the Royal Astronomical Society is a copy of Brewster's original in her handwriting.

[12] The copy of Brewster's letter claims the original was sent on 19 January, whereas Erskine's scribe claims it was sent on 15 January. Whatever date is incorrect, one can bet that very little time flowed between the reception of the letter by Erskine and its transmission to Herschel.

But there is no trace of subsequent correspondence between Thomas Young and David Brewster before 1815 (Brewster 1815), by which time the latter would try and introduce Young's law of interference in his own projectile theory of light (Brewster 1818: 271-272).

Which finally leads us to enquire the possibility that William Herschel would have personally communicated the content of that letter to Thomas Young. An enquiry that cannot be totally disentangled probably from another one, focused on the reason why Herschel did not transmit that letter in the first hand to the Royal Society.

On that matter, what remains from the correspondence carried by Brewster and Herschel at the time is of little help. It actually starts with a letter from Herschel sent on March 1803 (Herschel 1803) – that is more than a year after Brewster sent him his paper (Brewster 1802a) and less than a month after Erskine urged Herschel to send it back to its author (Erskine 1803) – and seems to be triggered by another letter sent two months earlier by Brewster, which is missing today. There, Herschel finally states that he read Brewster's paper with great attention as soon as he had received it, more than a year before, as it is 'on a subject that has always been interesting to me'. Yet, he inserted there a number of 'pencil numbers', which referred to 'things that appeared to want illustration or confirmation, or seemed to be such as I could not admit' and that he briefly listed on a separate sheet of paper, for later development.[13] Yet, Herschel says, those annotations finally 'amounted to near 50' so that he could not find the time to extend his own ideas on those points and return them to Brewster in due time. Then, he promises he will try to develop on a couple of objections only, 'within a few days'.

Brewster's answer comes without delay (Brewster 1803). On the 16$^{th}$ of the same month, he expresses his gratitude for the time Herschel spent on his account, and his great 'satisfaction' in the case he would receive Herschel's sentiments on any part of his paper. Furthermore, Brewster lets the reader understand that he actually received the original of his 'observations on inflexion' back, with the 'pencil marks' on it, which he considers as 'extremely serviceable in directing my attention to those passages to which they are affixed'. After what he politely tries to get Herschel's precise opinion 'on my method of deducing the law of the inflecting force from the hyperbola figures of the fringes', but obviously cannot proceed much further by himself. So that he finally switches to a completely different topic, concerning the best way of producing metallic mirrors for telescopes.

The next letter we have already dates 7 January 1805 and is signed Brewster again (Brewster 1805). Its introduction suggests that he received one other letter from Herschel in the meantime, arrived a long time before. And that this letter included at least some of the remarks that were promised; which were made with such 'frankness' that Brewster fears Herschel could have interpreted his long silence as a sign of 'dissatisfaction'. He then tries and justify his position on a couple of early points in his paper that are of no matter to us, before acknowledging Herschel's answer on the problem of the hyperbolic fringes, which, 'if true, would certainly overthrow the reasoning in my paper'. Yet, he insists, that same 'hyperbolism of the fringes' – that we saw him oppose to Young's theory of inflection earlier (Brewster 1802b: 64) – actually 'cannot be accounted for on Newton's supposition, that the repulsion of the inflecting body commences where its attraction ceases', as he had already stated in his paper (Brewster 1802a). 'But that phenomenon may be explained by supposing attraction to begin where repulsion ceases; and even without the aid of a repulsive force'; which statement certainly sums up by itself the inner core of the disagreement between Brewster and Herschel on inflection – although it is never expressed as such. And which most probably explains why Herschel never transmitted Brewster's 'observations on inflexion' to the Royal Society: as one saw earlier, Brewster claims that a sole inflective force (directed towards the body) can justify for inflection. But as one may see later, Herschel is deeply dedicated to a model of inflection lying on the exertion on the projectiles of light of the combination of a 'deflective' force pushing them away from the material body, and an 'inflective' force exerted at a longer range, and probably decreasing as a lower power of the distance (Herschel 6/13).[14] But let us leave this point aside until we have explored the content of Herschel's last letter.

---

[13] Those numbers do appear on the copy of Brewster's paper on inflection that is kept at the Royal Astronomical Society, although the remarks themselves are now missing (Brewster 1802a).

As Brewster concluded the matter on inflection in his last letter by the affirmation that 'I would wish much to see your experiments on the inflexion of light. The subject is so new and interesting, that I think you should not withhold them from the public' (Brewster 1805), Herschel replies, three weeks later, that 'my remarks on the fringes of inflected light are founded on experiments that cannot be confuted, and will certainly (as you wish) be laid before the public' (Herschel 1805). A declaration which definitely closes the exchange those two great men had opened three years earlier on inflection.[15] However, as the examination of these letters did not offer much material, neither on the exact reason why Herschel did not transmit the initial letter to the Royal Society, nor on his own opinion on inflection, that declaration opens for us a new line of investigation, pointing towards the remains of Herschel's own works on inflection – or rather 'flection', as he would certainly specify it – of the rays of light.

**William Herschel's own works on flection**

Indeed, even though William Herschel finally never published any of his experiments on 'flection',[16] documents mentioning that phenomenon held in his Archives at the Royal Astronomical Society are numerous and multifarious. Nine files at least contain pages specifically concerning flection (Herschel 6/8; 6/9; 6/10.1; 6/12.1; 6/12.2; 6/13; 6/14; 6/15; 6/16).

An index of the experiments on that matter that he led from 1 January 1807 to 26 June 1808 – that is after the correspondence he pursued with Brewster on that matter – sets a list of 159 experimental manipulations (Herschel 6/14), which description covers some thirty-nine pages of a lab book (Herschel 6/10.1); and whose data are extensively collected on twenty-seven pages of another notebook (Herschel 6/10.2). No experiment prior to that period can be found – except one dating 8 September 1805 (Herschel 6/8) – though a handful of additional experiments are described on loose undated folios (Herschel 6/12.1; 6/16). However, a 'memorandum' summarizes more than forty 'experiments to be made' that directly concern inflection – most of which are not repeated in 1807 – making it probably a preparatory document for Herschel's long lasting experimental investigation on flection (Herschel 6/9). Three files gather a dozen different figures (plus some that are repeated form one file to the other) illustrating the course of light rays passing through a circular hole or a slit – or passing by an opaque disk or the edge of a rectilinear obstacle – and trying to assign the shape of shadows or penumbra that can be observed at different distances behind them to the trajectory of those rays (Herschel 6/13; 6/15; 6/16). Two versions of an essay entitled 'enumeration of the various modifications of the rays of light' try to identify and isolate as exhaustively as possible the fifteen types of modifications that light will suffer when passing through a transparent body (Herschel 6/10.1, 353-368; 6/12.1), and abruptly stops after the seventh type of modification it may suffer from opaque ones (Herschel 6/12.1); making those documents an extensive set of (re-)definitions of optical phenomena, including 'inflection' and 'deflection'. Finally, two pairs of two folios 'on the modifications of the flections of the rays of light' (Herschel 6/10.1; 6/12.1) can be interpreted as late attempts from Herschel to finally write up a synthesis of his investigations on flection.[17]

---

[14] Further reference will be made to documents located in the Herschel Archive at the Royal Astronomical Society of London. As most of them are not dated, we propose to refer them as we did here, by addressing the index they were given in those archives.

[15] By the way, in that last letter Herschel also declares he 'misplaced the memorandum I had taken of my remarks I had the liberty of sending to you' (Herschel 1805); coincidentally providing us with an explanation for the absence of those remarks as adjoined to the copy of Brewster's paper that was withheld by the Herschels (Brewster 1802a).

[16] At the exception of one experiment that he conceded to publish at the end of his 1807 article on coloured rings, as he felt it illustrated both phenomena (Herschel 1807a: 231-232).

Those two pairs of folios actually synthesize his position pretty clearly: light rays can suffer two kinds of deviations – or flections – as they pass by the edges of solid bodies. One towards the body, that he calls inflection. One outward, he calls deflection.[18] Such that Herschel's opinion – as confirmed by the whole set of those documents – clearly is that solid bodies exert not only an attractive force on particles of light – which, according to him, was the dominant opinion at the time (Herschel 1807b) and which was, as we saw, the case Brewster was trying to prove – but also a repulsive one that is more sensible at a shorter distance from the body. However, Herschel officially balks at 'framing hypotheses' (Herschel 6/12.1) and then does not elude the possibility that flection could instead be due to a power emanating from the rays themselves, of from 'some subtile (sic) atmosphere surrounding the body' (Herschel 6/10.1). Suggesting that he could well have been aware of Young's hypothesis.[19]

And just as important with respect to our present investigation, Herschel's 'memorandum of experim[ents] to be made relating to the powers of att[raction] and rep[ulsion]', sets a list of dozens of optical experiments, such as:
> 1/ To examine whether rep[ulsion] of deflect[ion] is a general property of all bodies, by a ray of light.
> 2/ Whether the compactness of bod[y] makes any change in the force of this power.
> 3/ Whether the polished surface makes any difference.
> 4/ Whether a long surface parallel to a ray will show different effects.

---

[17] On top of that, a few letters from Herschel's correspondence briefly discuss the problem of flection, as for instance a set of letters Herschel sent to Patrick Wilson in 1807 (Herschel 1807b; 1807c; 1807d).
[18] Besides those inflected and deflected rays, Herschel's papers seem to explore the existence of a third kind of rays, that he defines as 'black making' for their apparent property of making the direct surrounding of the shadow blacker than the shadow itself (Herschel 6/10.1; 6/13; 6/15; 6/16). But as they seem to appear at a later stage of his investigations and not to play a significant role in our present investigation, we will leave those 'black making' rays aside.
[19] This elusive reference to atmospheres as being the potential cause for flection might also be pointing to the works of Mairan (1740) and Du Tour (1768; 1774) – whose names were explicitly associated to the hypothesis of 'an atmosphere of a variable or uniform density surround[ing] all bodies' by Brewster (1802a: 1) – even though they are never mentioned in Herschel's documents on flection we collected.

> 5/ To examine whether a large heavy body approaching to a passing ray will show the effects of gravity. (Herschel 6/9: 1)

That is a series of experiments very clearly designed to explore the possible influence of the nature of the inflecting body on flection. Just like many other experiments from that list, actually comparing the 'flecting power of metals and cork or at least such light objects as have a polish, straw, glass, hair' (exp. 82), of 'unctuous obstacles' (exp. 83), of a 'knife […] dipt into boiling water' (exp. 87), of 'natural fine points' such as the 'sting of a wasp' or 'the thorns of a moss rose' (exp. 88), or of 'a very sharp, fresh-set raisor' (exp. 91). This document – which, by its overwhelming exhaustivity, exhibits a little further Herschel's inclination to a radical inductive methodology inherited from Bacon – is undated, but certainly dates much earlier than 1807.

For the long series of experiments on flection Herschel held along the first half of year 1807 never mentions this type of problems anymore (Herschel 6/10.1, exp. 692-929). Obviously, some of them involve an 'ivory ball' (exp. 692), a 'brass disk' (exp. 697), a 'slip of iron' (exp. 699), a 'large plate of lead' (exp. 783), a 'silver wire' (exp. 887), or 'globular grains of Jamaica pepper' (exp. 889). But never the nature of the screen, or inflecting body, seems to be examined in a systematic or comparative way. And most of the time, it is not indicated at all; whereas its shape and dimensions are always stated. Even clearer is the other experimental notebook on flection, compiling the data only from those same experiments, and which never states the nature of the body at all (Herschel 6/10.1) – confirming that in the other document this information was contextual only. Consequently, regarding the question on the relative power of 'different inflecting substances' that was, let us remind it, the question that Young presented as decisive with respect to his ether distribution hypothesis (Young 1802b: XVI, 17r) – and that is a question that, according to us, was first experimentally solved out by David Brewster (1802a) – it seems that at some point Herschel had asked himself that question with very much acuity and emergency (Herschel 6/9), but that he had finally come to the same conclusion that 'the breadth of the fringes, and their distance from the Inflecting body, [a]re universally the same' (Brewster 1802a:9). For, following Carlo Ginzburg's lead in interpreting the minute gestures and tracks people produce and leave unconsciously as a significant signature of their actual practice and intention (Ginzburg 1979), we want to interpret Herschel's absolute lack of interest for the nature of the inflecting body in his 1807 long and diversified series of experiments on flection as a proof of his knowledge of the answer to that question. Or, more precisely, of his certitude that the nature of the flecting body just does not affect the result of flection.

Whether Herschel discovered it by himself, or became aware of that fact reading Brewster's 'observations on inflexion' will probably stay a mystery. The brief passage of Brewster's letter relating his experiments on that matter bears eight of Herschel's pencil marks (Brewster 1802a: 8-9, marks 34-41), proving that he was sensitive to that matter. But was Herschel mainly refuting Brewster's attribution of the inflecting force to the ultimate particles of matter (marks 37; 41)? Or was he denying, approving, or maybe claiming priority for the discovery that 'the inflecting force does not increase with the quantity of matter in the inflecting body' (marks 34; 38-40)? That is something one could not find out and that would probably be ascertained only by the discovery of those sheets of paper containing Herschel's remarks as he sent them back to Brewster. Yet, in any case, considering the content of the numerous experiments on flection described in his archives, and considering his declaration to Brewster that he had performed many more before (Herschel 1805), it is pretty clear that Herschel had that knowledge by 1807. And there is little doubt that after reading Brewster's essay on flection at the beginning of 1802 he had it too; whether that essay brought him that information, confirmed previous experiments of his, or triggered a new set of experiments that led him to that same conclusion.

**Connecting Herschel – and thus Brewster – to Young**

As obliged as we feel to apologize for having made such a long detour via the examination of Herschel's private works on flection, we cannot help but feeling that it was worth the trip. For it allowed us uncover the fact that by that first half of year 1802, not only Brewster and Young, but Herschel too, were all fairly interested in inflection. That all three were defending competing theories of the phenomenon. And that all of them were finally convinced at that point that it would not depend on the inflecting substance. Such that we now detain more relevant matter to investigate Young's relation to Herschel at that period, and how this relation can have influenced Young's interpretation of inflection.

Indeed, Herschel and Young had known each other for a long time, as Thomas had already paid a visit to the astronomer at Slough as early as 1794, while still following his medical studies in London (Peacock 1855: 44). Still, as revealed by an examination of Herschel's archived and published papers from the beginning of the 19th century, he seems to 'evince not the slightest interest in Young's works' (Cantor 1983: 141). Indeed, the Herschel Archive apparently does not contain a single trace of Young's name and opinions; neither in Herschel's documents where the works of other authors on flection is questioned (Herschel 6/12.1; 1807c), nor when an astronomical observation of Saturn's rings that would definitely prove wrong the vibratory theory recently revived by Thomas Young is discussed with John Robison (1804a; 1804b) and Patrick Wilson (Herschel 1807b). More eloquent even is the total absence of reference to Young's works in Herschel's 1807 article on the 'coloured concentric rings, discovered by Sir Isaac Newton' (Herschel 1807a). A silence that is contrasting with the fact that Herschel could not be unaware of Young's new *Theory of Light*, in which his discovery of invisible calorific rays, situated beyond the red-end of the spectrum (Herschel 1800a), was cited twice, as an additional proof of the common vibratory nature of light and heat (Young 1802a: 33; 47).

Whether Herschel knew the exact details of Thomas Young's *Theory of Light* when he received and read Brewster's letter on inflection cannot be ascertained, as the *Theory* – although it was read to the Society on 12 and 19 November 1801 – was only published in the Philosophical Transactions of the following year. In fact, the minute books of the meetings of the Royal Society do not specify the name of its members present at those meetings; only those of the speakers obviously, of the 'strangers' to the Society that were occasionally invited by members, and of those members who invited them indeed. Such that on this basis, one cannot know whether Herschel was present or not at the meetings of the Society when Young's *Theory* was presented (Royal Society of London 1799: 357-373). However, it is ascertained that Herschel had invited 'Mr. Wilson' to the session held on 16 January 1800 (Royal Society of London 1799: 46). Such that he must have been there himself, when Young's opinion on the analogy between sound and light was first defended that day (Young 1800). As well as his first explanation of the deviations of light by the spatial distribution of ether; his attribution of inflection towards solid body to a decreasing atmosphere of ether surrounding them; and, consequently, his willingness to confront this opinion to 'comparative experiments […] on the inflection of light by substances possessed of different refractive powers' (Young 1800: 130).

Then, whether or not Herschel knew the result of such an experiment in January 1800, he probably did not communicate it to Young at that period, since the urge for that experiment to be made is repeated in Young's 1801 Bakerian Lecture (Young 1802a: 43). Yet, on the very occasion of this Lecture, the vibratory hypothesis certainly acquired a degree of dangerousness for the projectile theories it had never reached before. And one can guess that a natural philosopher as dedicated to the projectile model as Herschel was would not have missed such an opportunity to weaken the vibratory position. All the more so as he was then working on his own theory of inflection; a theory he hardly appreciated to see challenged – judging by the way he received Brewster's proposal. And although we noted that there is no written trace of him seizing this opportunity, Herschel had plenty of occasions doing it privately, possibly orally, as him and Young, as members of the Royal Society, were meant to meet on a weekly basis at the meetings of that institution. Of course, the presence of members to such meetings was not mandatory and, as we said, the names of the present members were not recorded in the minutes of those meetings; such that we cannot determine at which meetings they met with certainty. But there is no doubt it happened several times during year 1801. And again during the first half of year 1802; that is during the lapse of time running from the presentation of Young's Bakerian Lecture in November 1801 – rapidly followed by Herschel's reception of Brewster's letter on inflection –, to the withdrawal of the ether distribution hypothesis in Young's last Lecture at the Royal Institution, dated 17 May, or maybe to the presentation of the next of Young's paper at the Royal Society, on 1 July 1802.[20]

For that date might mark a last turning point in our story. That day, Thomas Young and William Herschel both read papers at the Royal Society in London, and as such, it is the only day of that whole period on which one can ascertain they were both present there. Herschel's paper concerned his own catalogue of nebulae, a topic that seems loosely connected to our present problem (Herschel 1802). But Young's was a complement to his vibratory *Theory of Light*, accounting for 'some cases of the production of colours, not hitherto described'; including that of the 'colours of fibres', partly attributed to the inflection of light around them (Young 1802c). Furthermore, a week later, Young was at the Royal Society again, to read a letter from Swiss natural philosopher Pierre Prévost (1802), explicitly built – from its very first words – on inferences from William Herschel's experiments on the transmission of heat through different transparent media (Herschel 1800).[21] An occasion on which there is little doubt that Herschel himself would have been present too. Hence, although there is no documentary evidence of a discussion between Herschel and Young on any of those two occasions, all conditions seem set for it to happen. And as Prévost's paper, read by Young, was questioning Herschel's precautions and conclusions on his experiments on transmitted heat – and concluding, by the way, that they did not ascertain any legitimate conclusion on the difference or identity of light and heat[22] – both were most certainly exposed to debate on that topic.

---

[20] All along that period, the minutes of the Royal Society offer no evidence that both Young and Herschel were present at a same meeting, but that of 1 July 1802. However, because they were reading a letter, inviting guests or offering presents, it is acted that Herschel was present at least on five of the thirty meetings that occurred from 12 November 1801 to 8 July 1802, while Young was present at twelve of them at least (Royal Society of London 1799: 357-573); testifying for their regular attendance to those events.

[21] Although it is marked as having been read on 1 July 1802 in volume 92 on the *Philosophical Transactions* – and filed there in-between the above-mentioned letters of Thomas Young and William Herschel read that same day – Thomas Young himself claims in the proceedings of the Royal Society he used to transcribe in the *Journals of the Royal Institution*, that he read it a week later, on the evening of 8 July 1802 (Young 1802d: 193-194). That claim is confirmed by the minutes for these sessions (Royal Society of London 1799: 562-573).

[22] 'Qu'en appliquant ces principes aux expériences de M. Herschel, l'appréciation devient plus exacte, mais dépend néanmoins de quelques circonstances accessoires, et jusqu'ici indéterminées. Que, dans ces mêmes expériences, la différence apparente entre l'interception de la chaleur et celle de la lumière, par les mêmes matières, n'établit aucune conclusion légitime sur la différence ou l'identité de la lumière et de la chaleur.' (Prévost 1802: 447)

Even more interesting, in the *Journals of the Royal Institution*, Thomas Young used to propose his own *Proceedings of the Royal Society*; in which the summary of Prévost's letter concludes with such words:

> Mr. Prevost observes that, this theory would be equally applicable to the opinion of those who consider heat as consisting in the undulations of an elastic medium; although he thinks this opinion liable to many objections, especially on account of the resistance which the motions of the planets must suffer from it. In a note added by Dr. Young, who communicated the paper, the assertion of Newton, is quoted in answer to this objection, yet Dr. Young confesses that Newton appears to have calculated erroneously: but he observes that if the slightest difficulty of this kind should occur from astronomical considerations, it might be avoided by considering the luminiferous ether as unconcerned in the phenomena of cohesion, and then its rarity might be assumed as great as we chose to make it. (Young 1802d: 194)

And this quotation happens to close a loop that was opened some pages earlier. Indeed, it first highlights a very direct link from Prévost's (and Herschel's) works discussed that day, to Young's vibratory theory of light. And more specifically to his ether distribution hypothesis, which actually raises questions about the possible free flow of ether through material body, and of 'considering the luminiferous ether as unconcerned [or not] in the phenomena of cohesion'. This argument precisely led Cantor to the conclusion that the presentation of Prévost's letter triggered a line of thought in Young's mind that played the decisive role in the abandonment of his hope in a theory of a universal ether that could justify simultaneously for optical, calorific, electrical, magnetic and cohesive phenomena; and, consequently, in the withdrawal of his ether distribution hypothesis.

Yet do we think that the public reading by Thomas Young, in the presence of William Herschel, of Prévost's letter directly concerning his work and indirectly – but manifestly – connected to the vibratory theory of light and to the question of the interaction between ether and matter, is also very likely to have been the occasion of a discussion on flection between them. Such discussion that can have led to a description by Herschel to Young of experiments proving the independence on the inflection of the nature of the inflecting body, such as those experiments on opaque and refractive bodies related in Brewster's letter. Of course, this hypothesis would not account satisfyingly for Young's certitude on the independence of inflection on the refractive power of bodies displayed in the pages of his notebooks dating for the lecture he gave at the Royal Institution on 18 May 1802; but it would still stand for a potential addition of these lines at the end of that page on the occasion of the second and last presentation of this same Lecture in May 1803.

After this last consideration on what could have happened on July 1802, one obviously cannot help feeling like that man, at night, looking for his glasses under a streetlight although he knows he lost them on the other side of the road; just because there is no light over there. Yet, – besides feeling partly relieved by the intuition that such is quite often the fate of the historian – one must stress that what happened on 1 or 8 July 1802 is not the problem at stake here. Indeed, the problem we tried to tackle so far was that of the identification of optical arguments that could have contributed to Young's withdrawal of his ether distribution hypothesis. And our own hypothesis, that we tried to put to the test, is that Brewster's 'observations on inflexion', that were sent to William Herschel on 19 January 1802 in order to be read in front of the Royal Society – but which were kept for more than a year in Herschel's hands before being returned to sender – could have been a reason of that withdrawal; since they explicitly contained negative results to the very experiment Young had long claimed to be able to invalidate the cause he then attributed to inflection. Although one obviously offered no factual proof of this conjecture, a substantial number of clues and traces were gathered that sustain it, and certainly make it more probable at the end than it was in the beginning. Such that there is little doubt, in the end, that William Herschel had convinced Thomas Young by the first half of year 1802 that the independence of the flection of light on the nature of the inflecting body had been proved experimentally. Whether he had done it on the basis of his own experiments, or mentioning those described by Brewster, and on what occasion he did it exactly, is yet to be discovered. Even though the eventuality that he never did so, and that Young heard of such experiments through a different channel still remains; just as well as the eventuality that he never heard of Brewster's experiments on the inflection of light at all, and gave up on his ether distribution hypothesis as a consequence only of the astronomical and mechanical considerations set forth by Professor Cantor. Yet, precisely because these considerations dissolved Young's hoped-for universal ether into a series of individual ones unconcerned with each other, were not we entitled to expect the mobilization of some strictly optical arguments at least, in order to finally justify the abandonment of those ether atmospheres that were likely to be at the origin of the inflection of light?

Indeed, the ether distribution hypothesis presented an undeniable robustness and plausibility within the strict boundaries of Young's optical system. Both because it inserted inflection into a coherent explanatory scheme with refraction and reflections (Young 1800: 127-128), and because it was compatible with Young's explanation of the periodicity of the inflection fringes on the basis of the law of the interferences of light (Young 1802a: 42-44). Such that a whole set of good reasons were certainly necessary in order to consider and finally accept its abandonment. Without a doubt, some of them were those arguments, external to optics, and pointing to the impossibility of considering a single and universal ether accounting for a unified system of natural philosophy, that were put forward by Cantor. Yet, as Thomas Young's notebooks suggest – and as our investigation tended to prove – the experimental proof of the independence of inflection of light on the nature of the inflecting body certainly delivered a fatal blow. However, as there was no absolute logical necessity to discard that hypothesis on the single result of that experiment,[23] the fact that Young already had another possible explanation of inflection at hand at the time is certainly not to be neglected either.

As a matter of fact, the mention of Hooke's hypothesis that the cause of inflection could just be the natural tendency of light to diverge in the 1801 Bakerian Conference is a proof that Young was already considering it as a possible alternative to his atmospheres at the time (Young 1802a: 43). And the swiftness with which Hooke's hypothesis seems to replace the ethereal atmospheres in the optical notebooks is another sign that it was already much plausible to Young beforehand (1802b: XVI, 17r). Yet, in the first time at least, this option had probably been lacking coherence with the rest of Young's optics: as a matter of fact, Professor Darrigol demonstrated that Hooke's hypothesis was already at stake in Young's argumentation against the classical objection that light could not be the result of vibrations of the ether, for otherwise light, like sound and water waves, would skirt around obstacles (Darrigol 2012: 176-179). But Young's concern at that stage was certainly still more to convince his reader that there was no detectable light in the shadow, rather than to account for the reason why light was there indeed. And there is good reason to project that he still needed to observe the 'colours of fibres' – that were the luminous fringes 'not hitherto described' that he had discovered inside the shadow of fibres of wool or silk little before 17 May 1802 (Young 1802b: XVI, 16v) and related to the

---

[23] That experimental result could well have been interpreted as the mere demonstration that uniform or varying atmospheres of ether presenting higher densities than that in air were indeed accumulated around solid bodies, but that their density and extent were depending on the nature of the surrounding medium only, for instance.

Society on 1 July that same year, as we just saw (Young 1802c) – in order to elaborate a definitive connection between the sets of bright fringes he observed outside and inside the shadows. So as to finally gather those two phenomena he initially categorized separately as 'colours by inflexion' and 'colours of fibres' under the same and single concept of 'diffraction', mechanically explained by the same 'natural' diverging behaviour of light waves in 1803 only (Young 1804: 2; 6). For as a matter of fact, history of sciences is rich of examples of hypotheses that were overthrown not only on the account of direct empirical objections, but also – and jointly – by the raise of new arguments in favor of alternative ones, and by the removal of objections against those.

**Epilogue**

One last remark might indirectly sustain our main hypothesis, though. Actually, Brewster eventually presented publicly some of his observations on inflection by summer 1817, including the experimental demonstration that the phenomenon does not depend on the nature of the inflecting body.[24] That is fifteen years after the works mentioned in the letter we have been talking about, but only a few months after the first publications on diffraction of Biot and Fresnel; both of which demonstrating separately – amongst other things – the independence of the inflection fringes on the inflecting substance.[25] Such that Nahum Kipnis suspected Brewster's late communication to be 'a priority claim and might have been a response either to Fresnel or Biot' (Kipnis 1991: 217). And in order to support his hypothesis and justify how Brewster could have been aware so early of any of those French experiments, Kipnis argued not only that Biot and Brewster knew each other and had been exchanging letters during this period. But also, that Brewster was corresponding with Thomas Young at the time, who had precisely received copies of Fresnel's memoir both from Fresnel and Arago in May and July 1816 (Fresnel 1816b; Arago 1816); such that Young had 'possibly sent one of them to Brewster'. And indeed, isn't it easy for us now to imagine how Thomas Young could have been struck by the evocation in Fresnel's first paper of those experiments that might have led to the withdrawal of his ether distribution hypothesis fifteen years earlier? Then, would not it be a plausible move from him to turn to Brewster – if he knew him to be the first to have run those same experiments – and tell him about it? Although this all remains, of course, a pure hypothesis.

In any case, that temporal coincidence of the late public reading in Edinburgh of Brewster's observations on inflection with the first publication of similar experiments in France – about which he could have been informed of, or not, by Thomas Young – is no conclusive evidence concerning our present investigation. But at least does the publication of such similar observations on inflection on three different occasions within the same year provide us with the definitive acknowledgement that, before 1816, the fact that its extent was independent on the nature of the inflecting body – that very argument that had probably sealed the fate of Young's ether distribution hypothesis as early as May or July 1802 – was still far from being common knowledge in the scientific community; such that it was certainly not unworthy investigating on the way Thomas Young had come aware of that matter.

---

[24] '*Proceedings of the Royal Society of Edinburgh*. JUNE 17th. Dr. Brewster laid before the Society several notices respecting light. […] From the experiments on inflexion, it follows that the deviation which the rays of light experience in passing by the edges of bodies, is not produced by any force inherent in the bodies themselves, but that it is a property of the light itself, and always appears, and has the same character whenever divergent light is interrupted by a shadow, whether the shadow is produced by cork, by platina, by diamond, by a dark groove in a metallic surface, or by a cylinder of flint glass immersed in a mixture of oil of cassia and oil of olives, that has precisely the same refractive power as the glass.' (Royal Society of Edinburgh 1817). Note that this last experiment, on inflection by flint glass immersed in a mix of oils equating its refraction index, offers a new and even stronger experimental argument against Young's ether distribution hypothesis.

[25] 'La nature du corps qui limite le milieu, ne change rien à cette loi.' (Biot and Pouillet 1816: 61) 'Le tranchant et le dos d'un rasoir, un fil métallique poli ou couvert de noir de fumée, et les corps dont les pouvoirs réfringens sont les plus différens, donnent toujours les mêmes franges.' (Fresnel 1816a: 244)

**Acknowledgements:** The author deeply thanks Professor Geoffrey N. Cantor for his warm reception of the thesis defended in this text and the ingenious and generous advice he gave for the improvement of a previous version of the article. One shall thank Professor Olivier Darrigol as well, for his very thorough comments preceding the editing process. Special thanks must also be addressed to Sian Prosser, librarian at the Royal Astronomical Society; Ruppert Baker and Virginia Mills, librarians at the Royal Society of London; Charlotte New, archivist at the Royal Institution; and Dan Mitchell, librarian at University College London, for their very kind and effective support. Last thanks should finally be addressed to Dr. Jean-Michel Manceau and Pb. Emma Gaucher for their decisive implication.

**Conflict of interest:** The author states that there is no conflict of interest.

**Data availability statement:** All the data mentioned in this paper are extracted from edited books and papers commonly available in the versions thoroughly cited in the following bibliography, and from manuscripts available for free consultation in public archives. More precisely, Thomas Young's notebooks are to be found in University College London (https://archives.ucl.ac.uk). The minutes of the meetings of the Royal Society are at disposal on request in the archives of the Royal Society of London (https://royalsociety.org/collections). And David Brewster's 'observations on flexion', as well as all his correspondence with William Herschel and David Steuart Eskine, and all the manuscripts on flection by William Herschel that were analyzed in this article are located in the Herschel Archives of the Royal Astronomical Society in London (https://ras.ac.uk/library), within the files whose references are quoted in the main text and precisely given in our bibliography.